\documentclass{article}
\usepackage{amssymb}
\usepackage{amsmath}
\usepackage{authblk}
\usepackage{bigints}
\usepackage{caption}
\usepackage{derivative}
\usepackage{enumitem}
\usepackage{etoolbox}
\usepackage{float}
\usepackage{gensymb}
\usepackage{geometry}
\usepackage{graphicx}
\usepackage{setspace}
\usepackage{tikz}
\usepackage{titlesec}
\usepackage{txfonts,xcolor}
\usepackage{comment}
\usepackage{verbatim}
\usepackage{url}
\usetikzlibrary{positioning, arrows.meta}

\date{08-01-2024}

\def\P{\mathbb{P}}
\def\g{\gamma}

\def\p{\phi}
\def\w{\omega}
\newcommand{\sqb}[1]{\left[#1\right]}
\newcommand{\ip}[1]{\left\langle #1 \right \rangle}
\newcommand{\pa}[1]{\left(#1\right)}

\def\om{\Omega}
\def \gt{\tilde{G}}

\title{Non-Gaussian statistics in static and dynamic Galton Boards}
\author[1]{Dhruv Shah}
\author[2]{R.K. Shishir}
\author[2]{Manjaree}
\author[2]{Shreya Pithva}
\author[2]{TY Booritth Balaji}
\author[2]{Rahul Agarwal Singh}
\affil[1]{Massachusetts Institute of Technology, Cambridge, Massachusetts 02139, USA}
\affil[2]{Indian Institute of Science, Bengaluru, 560012, India }
\date{}

\affil{}

\begin{document}

\maketitle

\begin{abstract}
  Perturbing the arrangements of pegs on a static Galton board can result in non-trivial stationary distributions, which in the continuum limit correspond to departure from regular gaussian behavior. Two such distributions are obtained. Further, the distributions generated for a ‘dynamic’ galton board under external forcing in  a general direction are obtained by solution of the corresponding stochastic differential equations. Exact cumulant generating functions for the distribution are presented for forcing in one dimension. An approximate expression, correct to first order in the forcing amplitude, is presented for the case of two dimensions. Both cases show nontrivial departures from the static gaussian solution.  
\end{abstract}

\section{Introduction}

A galton board \cite{Galtonboard} is a toy composed of balls falling vertically through a lattice of scatterers. Balls are released from the top, get scattered, and are allowed to fall into a set of adjacent bins. Originally developed to establish the central theorem, several theoretical and experimental studies have been done to modify the distribution of balls obtained in the bins. \\
Studies involve varying the geometry of scatterers, the properties of the scattering \cite{eres}, the shape of the scattered balls \cite{ngnd}  and the lattice structure. Experiment indicates that the no-slip galton board also obeys gaussian statistics. Certain exact calculations for the quantum generalisation of the board have also been studied. \cite{Hilbertgaltonboard} \\ 
In the continuum limit, the equations of motion for the ball through the static lattice can be converted to a stochastic differential equation with the appropriate noise statistics. The continuum limit of the galton board is essentially a non gaussian diffusion on a line. The fokker planck equation with a spatially varying diffusion coefficient has been studied \cite{vardif}. Methods used involve solving the SDE’s with multiplicative noise, with a diffusion coefficient of the form $D(x) \sim |x|^{-\theta /2}$. In section \ref{stationary} we obtain stationary solutions to the Fokker Planck equation for two special cases, namely $D(x) \sim (1 + ax^2)^{-1}$ and $D(x) \sim e^{-x}$.\\
The main problem studied in this work is the behaviour of the system under a driving force. Henceforth we shall call this system the ‘dynamic’ galton board. 
Only external time dependent forcing is considered, which would serve as a good model for mechanical forcing. Velocity-dependent forcing has been studied in various other works, such as the case of charged balls in a magnetic field\cite{magnet}.\\
External forcing in the 2D plane of the board produces some non-trivial, but intuitive departures from regular gaussian behavior. The coupled system of stochastic differential equations is solved to leading order in the forcing amplitude and higher order cumulants are computed analytically using standard methods. It turns out that the source of the non-gaussian behavior for a dynamic board is not the anomalous diffusion of individual balls, but, the varying initial conditions for different balls, due to which an overall non-gaussian spread is seen.

\section{Equations of motion for the Static Board}\label{Theo}

For a stationary board, all the balls are identical and it suffices to consider the statistics for a single particle. For the $j^{th}$ bin, the $N$ - ball probability is written as 
\begin{equation}
    P_N(j) \propto p_j
\end{equation}
, where $p_j$ is the probability for a single ball ending up in the $j^{th}$ bin.
In our calculations we consider totally inelastic scattering through the lattice of scatterers. Therefore the time for a ball to fall from one 'level' to another is given by a time-step ($\tau$), which is roughly constant for a static board, but depends on the $y$ coordinate for a dynamic board as shown in a later section. Note that the inelastic assumption is essential, for, in the case of elastic scattering it has been shown that at long times, the vertical displacement of the balls grows as $t^\frac{2}{3}$ \cite{Limit}. 
For each scatterer on the lattice with a horizontal co-ordinate $x_j = jl + x_0$, a ball is scattered to left(right) with probability $a_j$($b_j$), else it simply falls down. In general $a_i + b_i \ne 1$.
The master equation \cite{Master} is thus
\begin{equation} \label{Master}
    p_n(t + \tau) = b_{n-1}p_{n-1}(t) + a_{n+1}p_{n+1}(t)
\end{equation}

Coarse graining we get (for the static case), we get the usual fokker-planck equation \cite{Master}. Using biased pegs (say of different shapes, such that they preferentially scatter balls to one side), we vary the $c(x)$ term. Further, the lattice model can account for variations in peg density and the diffusion constant, by having 'holes', i.e. lattice sites $k$ such that $a_k = b_k = 0$. The equations of motion in the 2D plane can be represented by the set of stochastic differential equations: 
\begin{equation}
\begin{gathered}
    dx(t) = c(x) \, dt + \sqrt{2B(x)} \, dW(y) 
\\
dy(t) = \gamma g \, dt
\end{gathered}
\end{equation}

The wiener process is a function of $y$ and not $t$. This is because the scattering process occurs at intervals of constant vertical spacing. This is generally different from scattering at periodic time intervals, apart from when $y$ and $t$ are linearly related (static board).

The advantages of an entirely stochastic approach in finding analytical results are numerous. Nevertheless, direct kinematic approaches involving the individual collisions with scatterers may well be explored, such as those in \cite{lorentzgas},\cite{deterministic}.

\section{Some Non-Gaussian Stationary state solutions} \label{stationary}

For the static board, $t$ and $y$ are synonymous. We seek stationary solutions to to the fokker planck equation for nontrivial $c(x)$ and $D(x)$, which can be physically realized in terms of pegs as discussed before. Physically these 'stationary solutions' represent the steady distributions for boards of arbitrarily large height. 
The general formalism for obtaining stationary solutions to (3) is very well known \cite{Stocastic}: we want the probability current $j(x) = 0$, i.e. 
\begin{equation}
    c(x)p(x) + \dfrac{d}{dx} \Big(D(x)p(x)\Big) = 0
\end{equation}
Where $p(x)$ is the stationary distribution
This has the solution \cite{Stocastic}
\begin{equation}
    p(x) \propto \frac{e^{-\int dx \frac{c(x)}{D(x)}    }}{D(x)}
\end{equation}
Some very carefully chosen $c$’s and $D$’s yield stationary normalizable solutions, and we present them here with some statistics:

\subsection{Exponentially distributed scatterers}

Let $c(x) = c_0 > 0$ and $D(x) = D_0e^{-x}$. 
This gives us, (with $\rho_0 = \frac{c_0}{D_0}$ ),
$p(x) = \rho_0e^x \exp(- \rho_0 e^x)$.\\
So we get the cumulant generating function to be
\begin{equation}
C(k) = \ln \mathbb{E}[e^{kx}] = \ln(\Gamma(k+1)) - k \ln \rho_0
\end{equation}
The cumulants are :
\begin{equation*}
        \langle x \rangle_c = C'(0) = -(\gamma + \ln \rho),
        \langle x^2\rangle_c = C''(0) = \dfrac{\pi ^2}{6}
\end{equation*}
\begin{equation*}
 \langle x^n \rangle _{c} = (-1)^{n} \zeta_{n} \Gamma(n), \, \, n \geq 2 
\end{equation*}
Here $\gamma$ is the euler-mascheroni constant.

\subsection{Landau Distribution}

Let $D(x) = \dfrac{D_0}{l^2 + x^2}$ and $c(x) = c_0>0$. Using (7) this gives us
\begin{equation}
    p(x) = \dfrac{\rho_0}{2 \sinh{\left(\dfrac{\rho_0 \pi}{2l}\right)}} \dfrac{\exp\left(-\dfrac{\rho_0}{l} \arctan{\dfrac{x}{l}}\right)}{x^2 + l^2}
\end{equation}
The distribution is heavy tailed, and it is easily seen that $\langle x^2 \rangle$ and all higher moments diverge. This is a Levy alpha-stable curve \cite{Maths}, with $\alpha = 1$.\\ In fact, due to the asymmetry of the drift, even the mean diverges. This belongs to the``Landau'' class of distributions.\cite{Landau}

\section{The Dynamic Galton Board}
\subsection{Sinusoidal 1D Forcing}\label{Sine}

The theoretical description of particles diffusing in a vibrating board is done in the board’s frame of reference. In this frame, there is a periodic pseudo force due to vibration. Just as before in section 1, we will assume a zero-inertia motion, i.e. the acceleration of the particle in both horizontal and vertical direction on average is negligible due to constant collisions with the pegs. We assume a periodic forcing in the x direction of the form $F = F_0 \sin(\omega t )$
\\
The set of stochastic differential equations for this case is 
\begin{equation}
\begin{gathered}
    \mathrm{d}x(t) = \sqrt{2B}\ \mathrm{d}W(t) + \g f_0\sin(\omega t + \p)dt \\
    dy (t) = \gamma g dt
\end{gathered}
\end{equation}
    
where $\gamma$ is the damping. Note that $t$ is again synonymous with the variable for the height ($y$); as before we change $B$ to $D$ and use $t$ in place of $y$.
The crucial point that modifies the final distribution is that the balls are released at different times. So, the initial phase in the harmonic forcing that they will start with, $\p_0$ is different for different balls.
\\Assuming that the rate of balls released is constant in time, we have that for the $N$ balls ($N >> 1$), their starting phase angles are uniformly distributed.
\begin{equation*}
        q_0(\p) = \frac{1}{2\pi}; \ \quad
        \mathrm{d}x_{\p}(t) = \sqrt{2D} \  \mathrm{d}W(t) + \g f_0\sin(\omega t + \p)dt        
\end{equation*}

For the normal diffusive case ($D = \text{const}$), ignoring constants we get.
\begin{equation}
\begin{gathered}
    x_{\p}(t) = \sqrt{2D}W(t) + \frac{\g f_0}{\omega} \left[ \cos(\p) - \cos(\omega t + \p) \right]\\
\end{gathered}
\end{equation}

The fourier transform of the total probability distribution (for $N$ balls having fallen through the board) is (Details in appendix \ref{appendix:A})

\begin{equation}
    \mathbb{\widetilde{P}}_N(k, t) \propto J_0 \left( 2k \beta \sin \left( \frac{\w t}{2} \right)\right) e^{-Dk^2t}
\end{equation}
where, $J_0$ is the $0^{th}$ order Bessel function with $\beta = \frac{\g f_0}{\omega}$.

Taking the logarithm and using simple properties of bessel functions, the cumulants are easily calculated as
\begin{equation}\label{equation:length}
\begin{gathered}
    \langle x \rangle = 0 \, \, \, , \, \, \, \langle x^3 \rangle_c = 0\\
    \langle x^2 \rangle_c = 2Dt + 2v^2 \, \, \, , \, \, \,   \langle x^4 \rangle_c = 6v^4  \\
\end{gathered}
\end{equation}
Above, we have used $v = \beta \sin(\w t /2)$.
There is non gaussian behavior in the fact that $\langle x^4 \rangle_c \neq 0.$
As one may expect, the periodic forcing results in an increase in the variance, which also oscillates in time.\\

(\ref*{equation:length}) also suggests a length scale in the problem, i.e. $\beta = \frac{\g f_0}{\omega} $. The first term in the variance is $\sqrt{2Dt} = l\sqrt{\frac{t}{\tau}}$.
For the non-Gaussian effects to be prominent $\beta \gtrsim \sqrt{2Dt}, i.e., \frac{\g f_0}{\omega} \gtrsim l\sqrt{\frac{t}{\tau}}$.
Also, $\frac{t}{\tau} = \frac{H}{h} = \frac{H}{l}$, where h is the vertical lattice spacing, and H is the height through which the balls fall, (i.e. the height of the board). The above condition thus is $\g f_0 \gtrsim \omega \sqrt{lH}$

\subsection{Effect on an already non-Gaussian distribution}

Suppose we have a given distribution and shape of scatterers (i.e. some nontrivial $c(x)$ and $D(x)$) so that the equations of motion in absence of periodic forcing is :
\begin{equation}
    \begin{gathered}
           dx(t) = c(x) \, dt + \sqrt{2D(x)} \, dW(t) \\
    \end{gathered}
\end{equation}
Suppose the corresponding fokker planck equation for the probability density admits a solution $Q(x,t)$. Then in presence of a forcing, the equation of motion for a ball with released at a given phase $\p_0$ is 
\begin{equation}
dx(t) = c(x) \, dt + \sqrt{2D(x)} \, dW(t) + \g f_0 \sin(\omega t + \phi_0) \, dt
\end{equation}
The corresponding fokker-planck equation will simply have a shifted solution :
\begin{equation}
p(x,t \, |\, \p_0) = Q(x - \g f_0 \, \w^{-1}  \, ( \cos(\p_0) - \cos(\omega t + \p_0))\, , \, t)
\end{equation}
In $k$ - space there is only an extra multiplicative factor of $\exp \, (ik \beta \,(\cos(\p_0) - \cos(\omega t + \p_0)))$ . \\
Following the same procedure as in Appendix \ref{appendix:A} we get that the total distribution factorizes into the individual static (zero forcing) component, multiplied by a force-dependent component. Thus the cumulant generating function only has an extra (time dependent) additive term and is given by :
\begin{equation}
\begin{gathered}
    C_{forced}(k,t)  = \ln \widetilde{Q}(k,t) + \ln J_0  (2vk)  = C_{static}(k,t) + \ln J_0 (2vk)
\end{gathered}
\end{equation}
$\beta$ and $v \equiv v(t)$ have been defined in the previous section.\\
It follows that the effect of forcing in the $x$ direction only gives an additive contribution to the results of the previous section, to each of the cumulants. This is independent of the nature or distribution of scatterers.

\subsection{General horizontal periodic Forcing}
Consider a periodic external (zero-mean) forcing of period $\frac{2\pi}{\o}$. This can be expressed as 
\begin{equation}\label{thet}
    f(t) = \sum_{n = 1}^\infty f_n \cos{(n\om t + \theta_n)}
\end{equation}
The motion in $y$ is linear with time step $\tau = \frac{a\g}{g}$.\\

The initial condition for each ball is defined by the vector   $\boldsymbol{\Phi} = \left( \p _1 , \p _2 , \dots  \right)$. Here $\p _{j}$ denotes the phase of the $j^{th}$ sinusoidal component at the time of the release of the ball. Taking the large $N$ limit, with a sufficiently fast rate of release of balls, we can 
 assume as before that the initial phase angles of the balls are uniformly distributed, so that 
$$q_{l} (\p _{l}) = \frac{1}{2 \pi} \, \; \, \forall l$$

Integrating over the $\p _{l}$'s, we get
\begin{equation}\label{thetas}
    \begin{gathered}
    \widetilde{\mathbb{P}}(k,t)  \propto \int\, \prod_n     \left[  \frac{d\p_n}{2\pi} \exp{\pa{-ik\g \sum_{n - 1}^\infty \frac{f_n}{n\om}\sqb{\sin{(n\om t + \p_n)} - \sin{\p_n}}}}    \right]  \times e^{-Dk^2t}\\
    =e^{-Dk^2t} \prod_nJ_0\pa{\frac{2\g k f_n}{n\om}\sin{\pa{\frac{n\om t}{2}}}}\\
    \end{gathered}
\end{equation}
Note that the $\theta_n$'s in (\ref{thet}),(\ref{thetas}) are automatically absorbed into the $\phi _n$'s, and do not affect the fact that the initial phases of the balls are uniformly distributed.\\
The different fourier components again factorize, and the cumulant generating function as well as the cumulants are found to be :
\begin{gather}
    C(k,t) = -Dk^{2}t + \sum_n \ln{J_0\pa{\frac{2\g k f_n}{n\om}\sin{\pa{\frac{n\om t}{2}}}}}
\end{gather}
\begin{gather*}
    \ip{x} = 0\\
    \ip{x^2}_c = 2Dt + \frac{2\g^2}{\om^2}\sum_n \frac{f_n^2}{n^2}\sin^2{\pa{\frac{n\om t}{2}}}\\
\end{gather*}

The response for non-periodic forcing is a simple extension of the above. Writing $f(t) = \int_{-\infty}^\infty \frac{d \omega}{2\pi} e^{i\omega t}\widetilde{f}(\omega) $
,we get 
\begin{equation}
    \widetilde{p}(k, t \, \, | \, \, \Phi(\omega)) \propto \exp{\pa{ik\g \int_{-\infty}^\infty \frac{d\omega}{2\pi} \pa{\frac{e^{i\omega t} - 1}{\omega}} \widetilde{f}(\omega) e^{i\Phi(\omega)}}}e^{-Dk^2t}
\end{equation}

Here, $\widetilde{p}(k, t \, \, | \, \, \Phi(\omega))$ is the distribution for a single ball given the initial phases $\Phi (\omega)$. This is the natural extension of the vector $\boldsymbol{\Phi}$ defined after (\ref{thet}).\\

The many-particle limit is simply a functional integral given by 
\begin{equation}
    \widetilde{\P}(k, t) \propto e^{-Dk^2t}\int D\Phi(\w) \exp{\sqb{ik\g \int_{-\infty}^{\infty}\frac{d\w}{2\pi}\pa{\frac{e^{i\w t} - 1}{\w}} \widetilde{f}(\w)e^{i\Phi(\w)}}}
\end{equation}

\subsection{Response to 2D sinusoidal forcing}
Qualitatively, if we vibrate the board in both $x$ and $y$ direction with some frequencies $\w_1$ and $\w_2$, we expect different results as compared to the 1D case.\\
The reason is that displacements in $y$ are no longer proportional to $t$. We must construct $p(x, y)$, i.e. the distribution in $x$ for a height $y$ through which the balls fall, from scratch; since we can no longer claim that simple substitution of $y = \frac{at}{\tau}$ into $\P(k, t)$ yields the correct result. As we shall see, it does not.\\
The forcing is 
\begin{equation}
    \Bar{f}(t) = 
    \begin{pmatrix}
        f_1 \sin(\w_1t + \p)\\
        f_2 \cos(\w_2t + \psi)
    \end{pmatrix}    
\end{equation}
The above form of the forcing is for a single particle (ball) released with initial phases $\p, \psi$. In the many ball limit, the phases are distributed in $(-\pi , \pi]$ according to a density $\rho(\p, \psi)$. In the previous section, we assumed that the phase $\p$ was uniformly distributed. The equivalent condition here is $\int_{-\pi}^{\pi} d\p \, \, \rho(\p, \psi) = \int_{-\pi}^{\pi} d\psi \, \,  \rho(\p, \psi) = (2\pi)^{-1}$.
Another physical assumption we impose is that $\rho(\p, \psi) = G(\p - \psi)$ for some function $G$. We define $G(\theta) = (2 \pi)^{-1} \sum_{l} e^{-il\theta} \, \gt_{l}.$ 
The equations of motion in the overdamped limit are (writing $D'= \frac{D \tau}{a}$),
\begin{equation}\label{y vib y}
\begin{gathered}
    \text{d}x(t) = \g f_1 \sin{(\w_1t + \p)}  \ \text{d}t + \sqrt{2D'} \text{d}W(y)  \\
    \text{d}y(t) = \g f_2 \cos{(\w_2t + \psi)}  \ \text{d}t + \g g \text{d}t  
\end{gathered}
\end{equation}
As noted earlier, the argument for the wiener process is $y$ and not $t$. This is because the scattering of a peg at periodic intervals of $y$ (not $t$). The integral for the deterministic forcing is still in $t$, as the forcing is harmonic in $t$.\\
An exact solution is not feasible. We will work to first order in $f_2$.\\
Solving (\ref{y vib y}) for $y(t)$ we get
\begin{gather}
    y(t) = \g gt + \frac{\g f_2}{\w_2}\sqb{\sin{(\w_2t + \psi)} - \sin{\psi}}
\end{gather}
For $\frac{f_2}{w_2g} \ll 1$, $y(t)$ is invertible. We will use the inverse function $t(y)$ in (\ref{y vib y}) and write the SDE in $y$ as 
\begin{equation} \label{inverse vib}
    \text{d}x(y) = \g f_1 \, \, \text{d} \left[  \cos((\w_1t(y) + \p) \right]  + \sqrt{2D'} \text{d}W(y)
\end{equation}
the crucial part will be to wisely invert $y(t)$ into $t(y)$ and use it in (\ref{inverse vib}). \\
Consider a general relation of the form 
\begin{equation} \label{x + sin x}
    u = v + a\sin{v}
\end{equation}
We approximate the inverse as 
\begin{equation} \label{x - sin x}
    v \approx u - a \sin{u}
\end{equation}
with $a \ll 1$.\\
One can roughly estimate the deviation of this 'guessed' inverse against the true one for a suitably small $a$ and a suitable range of $u, v$ by following the procedure set in \cite{inverse}.
The motivation to choose a form $u - a \sin u$ is to have the same average slope in the relations (\ref{x + sin x}) and (\ref{x - sin x}) for $a \ll 1$ and to have periodic sinusoidal perturbations to the linear relations.

Using the same replacement of (\ref{x + sin x}) with (\ref{x - sin x}) for our case (with a few change of variables to account for constants) we get the replacement for $y(t)$ as 
\begin{equation} \label{inv}
    \begin{gathered}
    y(t) = \g gt + \frac{\g f_2}{\w_2}\sqb{\sin{(\w_2t + \psi)} - \sin{\psi}} \Longrightarrow \\
    t \approx \sqb{\frac{y}{\g g} + \frac{f_2}{\w_2g}\sin{\psi}} - \frac{f_2}{\w_2g} \sin{\pa{\frac{\w_2y}{\g g} + \psi}}
    \end{gathered}
\end{equation}
We can now consider the motion of a ball with initial phases $\p, \psi$. 
\begin{equation}
    \begin{gathered}
        \text{d}x(y) = \frac{\g f_1}{\w_1} \, \,  \text{d}\sqb{\cos{\pa{\w_1t + \p}}} + \text{d}W(y) \sqrt{2D'}
    \end{gathered}
\end{equation}
We will use subscripts $\p, \psi$ to remind us that the equations apply to a single trajectory with initial phases $\p, \psi$.
\begin{equation}
    \begin{gathered}
        \ip{x_{\p, \psi}(y)} = \frac{\g f_1}{\w_1} \, \, [\cos{\pa{\w_1t(y) + \p}}-\cos(\p)]\\
        \ip{x_{\p, \psi}(y)x_{\p, \psi}(y')} = 2D'\min{(y, y')}\\   
    \end{gathered}
\end{equation}
It follows that 
$$\widetilde{p}_{\p,\psi}(k, y) = e^{ik \ip{x_{\p, \psi}(y)}} \times e^{-D'k^2y}$$

Next we must average over the phases. The fourier transform of the total probability distribution is given by
\begin{equation}\label{ptil}
    \tilde{\P}(k, y) \propto e^{ - D'k^2y}\times \int_{-\pi}^{\pi} d\p \, d\psi \, G(\p - \psi) \, \, e^{ik \ip{x_{\p, \psi}(y)}}
 \end{equation}
We expand all terms to first order in the vertical forcing amplitude and compute the integrals. We note that in the above definition of the fourier amplitudes of G, we have $\gt_{l} = \gt^{*}_{-l}$. So, $\gt_{0} \in \mathbb{R}$, and we may write $\gt_{1} = he^{i \delta}, \, \, \gt_{-1} = he^{-i \delta}$ with $h, \delta \in \mathbb{R}$. Writing $\Gamma = h/\gt_0$, the cumulant generating function is (details in Appendix B) :
\begin{equation} \label{cgf2}
    C(k) = -D'k^2y + \ln \left( \gt_{0} \, J_0(kv) \right) -2ikd \Gamma \left[ sin\left( \delta + \left( \sigma_1 + \frac{\sigma_2}{2} \right)y \right) - \frac{J_2(kv)}{J_0(kv)} \sin \left( \delta + \frac{\sigma_2}{2}y \right) \right]
\end{equation}
 Here, $ \sigma_i = \w_i / \g g$, $ v = 2 \beta_1 \sin (\sigma_1 y / 2) $ and  $d = \frac{1}{2} \pi \sigma_1 \beta_2 \beta_1 \sin(\sigma_2 y /2)$, with $\beta_i = \g f_i / g$. $J_n$ is the bessel function of the $n^{th}$ order. Note that the expression is correct to first order in $\beta_2$.\\
It is now straightforward to compute the cumulants. They are :
\begin{equation} \label{cum2}
\begin{gathered}
        \ip{x}_c = \pi \Gamma \sigma_1 \beta_1 \beta_2 \sin \left( \frac{\sigma_2 y }{2}\right) \sin \left(\delta + \left(\sigma_1-\frac{\sigma_2}{2} \right)y \right)  \\
        \ip{x^2}_c = 2D'y + 2 \beta_1^2 \, \sin^2 \left( \frac{\sigma_1 y }{2}\right) \\
        \ip{x^3}_c = \frac{3}{4}\,\Gamma \pi \, \beta_1^3 \, \beta_2 \sin^2 \left( \frac{\sigma_1 y }{2} \right) [\cos(\delta) - \cos(\delta + \sigma_2 y)]\\
        \ip{x^4}_c = 6 \left[ \beta_1 \sin \left( \frac{\sigma_1 y }{2} \right) \right]^4
\end{gathered}
\end{equation}

Some comments are in order. Firstly, we see for $\w_2 = 0$, we get the results of 1D forcing as described in the previous section. It is evident from the cgf that all even moments are unchanged.\\
To first order in $\dfrac{f_2}{\w_2g}$, the mean and skewness are nonzero!\\ 

We give a simple visual explanation for this. For simplicity set $\w_1 = \w_2$, which corresponds to moving the galton board on an elliptical trajectory in the X-Y plane. Suppose in the first quarter of the cycle the board is moved right and down. The balls fall down slower relative to the moving board, which allows for a longer 'timestep' (For the ball to fall from one layer of pegs to another) a larger effective time for drift to one side. In another quarter of the cycle, when the board is moving upward, the timestep gets shortened. This induces the bias as seen in (\ref{cum2}). Notice that the $y$-averaged skewness is still zero. This is because, in alternate quarters (say first and third, or second and fourth) in the cycle the 'timestep lengthening' and 'timestep shortening' effects simply cancel out.

\section{Discussion}

In this work we have attempted to study non-Gaussian statistics generated by the Galton board. We have studied the system with the assumptions of inelastic, zero friction scattering off the pegs, and simply used a stochastic noise to model the collision process. In a regime where such assumptions hold good, the statistics of the Galton board are simply the solutions of the usual Fokker Planck equation. Another regime involves almost fluid-like no-slip scattering over large pegs, as studied in \cite{noslip}, \cite{vfriction}. \\

We have seen that for a dynamic galton board, horizontal forcing simply yields additive contributions to the cumulant generating function. Introducing 2D forcing is more interesting, for it induces a skewness in the distribution which varies periodically for a sinusoidal forcing. For a more general kind of (periodic) forcing, one qualitatively expects the same height-dependent skewness; nevertheless the calculations for this are not straightforward: the scheme we have used to approximate the inverse in (29) no longer holds. It is important to note that the Non Gaussian behavior in the Galton board is a collective effect caused only due to unequal phase angles of the individual Gaussian profiles. 

\bibliography{References}
\bibliographystyle{ieeetr}

\appendix
\numberwithin{equation}{section}
\section*{Appendix:}\label{appendix}

\section{Computation of the time-dependent distribution for 1D sinusoidal forcing}\label{appendix:A}
For a single ball which starts off with an initial phase $\p$ we have that:
\begin{equation*}
\begin{gathered}
    x_{\p}(t) = \sqrt{2D}W(t) + \frac{\g f_0}{\omega} (\cos(\p) - \cos(\omega t + \p))\\
    \langle x_{\p}(t)\rangle = \dfrac{f_0}{\omega}\cos(\omega t + \p) \, \, \, , \, \, \, 
    \langle x_{\p}(t)x_{\p}(t')\rangle_c = 2D \, \langle W(t)W(t')\rangle = 2D \min(t, t')
\end{gathered}
\end{equation*}

\begin{equation}
    \therefore p(x, t|\p) \propto \exp{\left( - \frac{(x - \langle x \rangle)^2}{2\langle x^2 \rangle_c}\right)} 
\end{equation}
(All higher averages corresponding to the Wiener process are zero)\\
Now, the total probability (after N particles have fallen) is
\begin{equation}
\begin{gathered}
    \mathbb{P}_N(x,t) = \sum_{i=1}^{N} p_i(x, t)
    \approx N\bigintss_{-\pi}^{\pi} \frac{\mathrm{d}\p}{2\pi} \, \frac{\exp {\left( - \frac{(x - \beta \cos(\p) + \beta\cos(\omega t + \p))^2}{4Dt}\right)}}{\sqrt{4\pi Dt}} \,\\
        \propto \int_{-\pi}^{\pi}\,d\p \int_{-\infty}^{\infty} dx \, e^{-ikx} e^{2ik\beta\sin(\omega t/2 + \p) \sin(\omega t/2)} e^{-Dk^2t}\, 
    \end{gathered}
\end{equation}
Using the integral definition of the bessel function, we get 
\begin{equation}
    \mathbb{\widetilde{P}}_N(k, t) \propto J_0(2kv)e^{-Dk^2t}\\
\end{equation}
The quantities $v \equiv v(t)$ and $\beta$ have been defined earlier.

\section{Computation of the cumulant generating function for 2D forcing} \label{appendix:B}

Using (\ref{inv}) we expand the exponential term in (\ref{ptil}) to first order in $f_2$. We get that \begin{equation}
    e^{ik \ip{x_{\p, \psi}(y)}} = e^{ ik \beta_1 (\cos(\p) - \cos(\p + \sigma_1y))} \times \left[ 1 + i \sigma_1 \beta_1 \beta_2 k \sin(\p + \sigma_1y)(\sin(\psi) - \sin(\psi + \sigma_2 y)) \right]
\end{equation}
Here, $ \sigma_i = \w_i / \g g$ and $\beta_i = \g f_i / g$.
Introducing the fourier expansion $G(\p - \psi) = \sum_{l} e^{il(\psi - \p)}\gt_{l}$ in \ref{ptil} and integrating over $\psi$ we get :
\begin{equation}
    \tilde{\P}(k, y) \propto e^{ - D'k^2y}\times \int_{-\pi}^{\pi} d\p \, \, e^{ik \beta_1 \Delta(\p)} \left[ \gt_0 + i \sigma_1 \beta_1 \beta_2 \sin(\p + \sigma_1 y) \sin\left(\frac{\sigma_2 y}{2}\right) \left( e^{i\p + i \sigma_2 y/2} \, \gt_{-1} +  e^{-i\p - i \sigma_2 y/2} \, \gt_{1}\right) \right]
\end{equation}
Here $\Delta(\p) = \cos(\p) - \cos(\p + \sigma_1 y)$.\\

Some algebraic manipulation along with the integral definitions of the bessel functions yields :
\begin{equation}
    \tilde{\P}(k, y) \propto e^{ - D'k^2y} \left[ J_0(kv) - kd J_0(kv)\left(  e^{-i(\sigma_1 - \sigma_2/2)y} \gt_{-1} - e^{i(\sigma_1 - \sigma_2/2)y} \gt_1 \right) - kdJ_2(kv)\left(  e^{-i\sigma_2 y/2} \gt_{-1} - e^{-i\sigma_2y/2} \gt_1 \right)  \right]\\
\end{equation}

The quantities $k , \, d$ are defined after (\ref{cgf2}).
Finally, we substitute $\gt_{1}/\gt_0 = \Gamma e^{i \delta}$ and $\gt_{-1}/\gt_0 = \Gamma e^{-i \delta} $. Taking the logarithm immediately yields (\ref{cgf2}).
\end{document}